\documentclass[12pt]{spieman}  
\usepackage{amsmath,amsfonts,amssymb}
\usepackage{graphicx}
\usepackage{setspace}
\usepackage{tocloft}

\title{A test platform for the detection and readout chain for the \textit{Athena} X-IFU}

\author[a,*]{Gabriele Betancourt-Martinez}
\author[a]{Fran\c{c}ois Pajot}
\author[a,b]{Sophie Beaumont}
\author[a]{Gilles Roudil}
\author[b]{Joseph Adams}
\author[c]{Hiroki Akamatsu}
\author[b]{Simon Bandler}
\author[a]{Bernard Bertrand}
\author[c]{Marcel Bruijn}
\author[a]{Florent Castellani}
\author[d]{Edoardo Cucchetti}
\author[e]{William Doriese}
\author[a]{Michel Dupieux}
\author[d]{Herv\'{e} Geoffray}
\author[c]{Luciano Gottardi}
\author[f]{Brian Jackson}
\author[f]{Jan van der Kuur}
\author[g]{Mikko Kiviranta}
\author[b]{Antoine Miniussi}
\author[d]{Phillipe Peille}
\author[c]{Kevin Ravensberg}
\author[a]{Laurent Ravera}
\author[e]{Carl Reintsema}
\author[b]{Kazuhiro Sakai}
\author[b]{Stephen Smith}
\author[b]{Nicholas Wakeham}
\author[c]{Henk van Weers}
\author[h]{Doreen Wernicke}
\author[b]{Michael Witthoeft}

\affil[a]{Institut de Recherche en Astrophysique et Plan\'etologie (IRAP)/CNRS, 9 Av. du Colonel Roche, 31400 Toulouse, France}
\affil[b]{NASA/Goddard Space Flight Center, 8800 Greenbelt Rd., Greenbelt, MD 20771, United States}
\affil[c]{SRON Netherlands Institute for Space Research, Niels Bohrweg 4, 2333 CA Leiden, The Netherlands}
\affil[d]{Centre National d'\'Etudes Spatiales (CNES), 18 Av. Edouard Belin, 31400 Toulouse, France}
\affil[e]{National Institute of Standards and Technology (NIST), Boulder, CO 80305, United States}
\affil[f]{SRON Netherlands Institute for Space Research, Landleven 12, 9747 AD Groningen, The Netherlands}
\affil[g]{VTT Technical Research Centre of Finland Ltd, Tekniikantie 21, Espoo, Otaniemi, Finland}
\affil[h]{Entropy GmbH, Gmunder Str. 37a, 81379 M{\"u}nchen, Germany}

\cftpagenumbersoff{figure}
\cftpagenumbersoff{table} 
\begin{document} 
\maketitle

\begin{abstract}
We present a test platform for the \textit{Athena} X-IFU detection chain, which will serve as the first demonstration of the representative end-to-end detection and readout chain for the X-IFU, using prototypes of the future flight electronics and currently available subsystems. This test bench, housed in a commercial two-stage ADR cryostat, includes a focal plane array placed at the 50\,mK cold stage of the ADR with a kilopixel array of transition-edge sensor microcalorimeter spectrometers and associated cold readout electronics. Prototype room temperature electronics for the X-IFU provide the readout, and will evolve over time to become more representative of the X-IFU mission baseline. The test bench yields critical feedback on subsystem designs and interfaces, in particular the warm readout electronics, and will provide an in-house detection system for continued testing and development of the warm readout electronics and for the validation of X-ray calibration sources. In this paper, we describe the test bench subsystems and design, characterization of the cryostat, and current status of the project. \end{abstract}


{\noindent \footnotesize\textbf{*}Gabriele Betancourt-Martinez,  \linkable{gabriele.betancourt@irap.omp.eu} }


\section{Introduction}
\label{sec:intro}  
The ESA-led \textit{Athena} mission \cite{2013arXiv1306.2307N} is poised to revolutionize our understanding of the hot and energetic universe. The $\sim$3000-pixel array of non-dispersive microcalorimeters on the X-IFU \cite{barret18} will provide imaging capabilities as well as wide-band, high-resolution spectra ($\sim$2.5\,eV FWHM up to 7\,keV). 
Technical progress towards mission milestones is excellent, but an important step that is yet to be done is a full test of the X-IFU baseline detection chain. NASA/Goddard Space Flight Center (GSFC) and the Netherlands Institute for Space Research (SRON) have working systems that pair GSFC kilopixel transition-edge sensor (TES) microcalorimeter spectrometers with custom-built room temperature electronics, but the mission baseline Digital Readout Electronics (DRE \cite{2018SPIE10699E..4VR}, led by the Research Institute for Astrophysics and Planetary Science [IRAP]) and Warm Front-End Electronics (WFEE \cite{2018SPIE10699E..4PC}, led by the Astroparticle and Cosmology Laboratory [APC]) have so far only been tested with X-ray pulses from simulated TESs. IRAP and the French National Space Agency (CNES), in collaboration with GSFC, SRON, and the National Institute of Standards and Technology (NIST-Boulder), is thus developing a test platform for the X-IFU detection chain. This will be the first demonstration of the representative end-to-end detection and readout chain for the X-IFU and an integral part of the preparation for the X-IFU flight instrument. The design, construction, and operation of the test bench will provide feedback on the detection chain subsystem designs, in particular the WFEE and the DRE, highlight any issues related to electronic interfaces between subsystems, and provide an in-house detection system for continued testing and development of the warm readout electronics and analysis software, and for the validation of future X-ray calibration sources. 

\section{Test Bench Subsystems}
\subsection{The Entropy GmbH cryostat} \label{ssec:cryostat}
The IRAP/CNES test bench cryostat (nicknamed ``Elsa") is an Entropy GmbH L-series Adiabatic Demagnetization Refrigerator (ADR). It has a pulse tube cooler and an ADR that consists of a superconducting magnet with magnetic shield and a double stage salt pill. The temperature stages are $\sim$35\,K, $\sim$3\,K, $\sim$500\,mK (GGG salt pill) and a base temperature of $\sim$35\,mK at the FAA salt pill stage. The pulse tube cooler from Sumitomo Heavy Industries provides 35\,W of cooling power at 45\,K and 0.9\,W at 4.2\,K, according to specifications. The cooldown procedure is fully automated and controlled by dedicated software, provided with the cryostat. The ADR recharge takes about one hour. All thermometry and housekeeping data are logged and can be displayed in real time. The thermometry provided with the cryostat is measured by an AVS 47 resistance bridge, and we installed an additional LakeShore 372 resistance bridge with eight thermometry channels, which is dedicated to cryostat characterization experiments and detector readout chain control. 

After installing a resistive heater dissipating 0.5\,$\mu$W through Joule heating on the FAA stage, we have measured a hold time of $\sim$11 hours while regulating the cold stage at 55\,mK. For comparison, the expected power dissipation for our detector system at the FAA stage (see Sec. \ref{ssec:fpa}) is $\sim$2\,nW, and $\sim$20\,$\mu$W at the 500\,mK stage\cite{gsfc_pc}. Our measured temperature stability while regulating the cold plate at 55\,mK is $\sim$3\,$\mu$K RMS. 

\subsection{Focal Plane Array and Cold Amplification} \label{ssec:fpa}

The Focal Plane Array (FPA) is situated at the coldest region of the test bench. Here, incident X-rays are absorbed by the detector's absorbers, their energy is measured by the TESs, and the signals from the TESs are read out and amplified using multiplexing electronics and superconducting amplifiers (Superconducting QUantum Interference Devices, or SQUIDs). The FPA is kept at 50~mK to optimize heat capacity and ensure good electro-thermal feedback, while staying below the superconducting transition temperature of the TES materials. 

\begin{figure}
\begin{center}
\begin{tabular}{c}
\includegraphics[height=10cm]{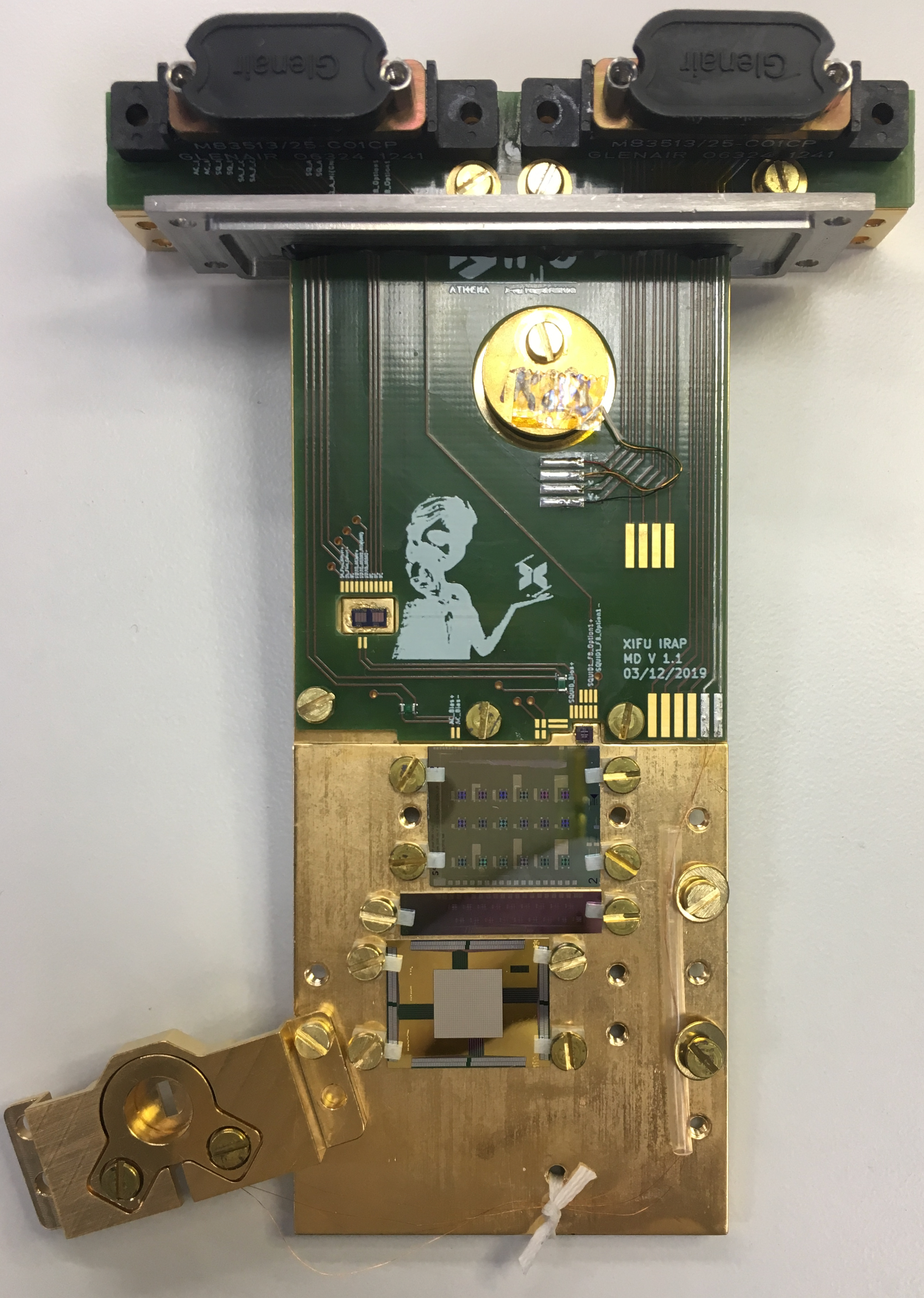}
\end{tabular}
\end{center}
\caption 
{ \label{fig:fdmfpa}
The IRAP/CNES FDM FPA. The field coil/collimator has been swung away from the plate to expose the TES. } 
\end{figure}

The original FPA design for the test bench was for Frequency Domain Multiplexing (FDM). The general design and electrical wiring for this FPA was adapted from the FDM FPA of the microcalorimeter group at NASA/GSFC, itself based on SRON designs\cite{2011ITAS...21..289D}, with slight optimizations for noise and the geometry of the Elsa cryostat. The completed IRAP/CNES FDM FPA is shown in Figure \ref{fig:fdmfpa}. It includes a 32x32 array of TES microcalorimeters from NASA/GSFC (``17.6 chip G") with 120\,$\mu$m TESs, no noise-mitigating stripes, a layer of copper on the back side of the chip for heatsinking, a transition temperature of $\sim$91\,mK, and a normal-state resistance of $\sim$30\,m$\Omega$. A $\sim$400-turn, $\sim$200\,Ohm field coil made from superconducting NbTi wire (0.06\,mm thickness) is placed over the TES chip. When current is passed through the wire, the coil applies a small magnetic field to the TES in order to null any residual external magnetic field present in the Niobium box that houses the FPA (described below). The field coil sits on a mechanical support that also serves as an X-ray collimator that only allows the active pixels to be illuminated, in order to reduce crosstalk. 

A FDM chip containing LC filters for the electrically connected pixels on the TES, and a transformer chip to perform impedance matching between the TES and SQUIDs, were provided by SRON. The filter chip has a common inductance of 2\,$\mu$H and varying capacitances to define the resonance frequencies, which are estimated to be between 900--4200\,kHz, with channels separated by 100-800\,kHz. The transformer chip has a turns ratio of 1:5. There are two SQUID amplifiers from VTT: a front-end SQUID that is connected to the LC filter chip and to the printed circuit board (PCB) mounted on the FPA, and a booster SQUID connected to the PCB. Our FDM FPA contains a K4-type front-end and a L6-type booster, which are 6-series 1-parallel and 32-series 4-parallel versions of the SQUID array design \cite{2015ITAS...2569234G}. These have been chosen to be compatible with each other and the parameters of the GSFC TES chip. The electrical signal from the TESs, transformer chip, LC filter, and SQUIDs are routed through a printed circuit board (PCB) with a design adapted from one developed at GSFC. The signals are passed to one of two MDM-21 connectors at the top of the FPA; one is for science (AC bias for TESs, SQUID bias, and SQUID feedback), and the other is for thermometry and housekeeping (field coil and thermometer). 

Wirebonds for electrical signals and heatsinking on the FPA were deposited with a wire bonding machine at the Laboratory for Analysis and Architecture of Systems (LAAS-CNRS). There are several sets of aluminum wirebonds for electrical signals: from the TES to the transformer chip, the transformer chip to the LC filter, the LC filter to the PCB and front-end SQUID, the front-end SQUID to the PCB, and the booster SQUID to the PCB. In addition, gold wirebonds for thermal heatsinking are placed at the corners of the TES chip and the front-end SQUID, connecting them to the gold-plated copper support plate underneath. 

Finally, a calibrated RuOX thermometer is mounted on the FPA for temperature measurement and control. The entire FPA is inserted into a Niobium magnetic shield which becomes superconducting below $\sim$9\,K. 

Electrical, mechanical, and optical compatibility between the FPA and both the GSFC and Elsa cryostats was ensured for ease of testing. The FDM FPA was transported to GSFC in December 2019 for initial tests. Resistance measurements at room temperature and the FPA cooldown were nominal. 

Just after the initial FDM FPA tests at GSFC, the \textit{Athena} baseline detector readout changed to Time Domain Multiplexing (TDM). The IRAP/CNES team is thus currently working closely with colleagues at GSFC and NIST to install an 8 column, 32 row TDM system into the Elsa cryostat. The initial installation will read out 2 columns. 

\subsection{X-ray Illumination and Filtering}

For standard measurement campaigns with the test bench, we will use a Fe-55 radioactive source from Eckert \& Zeigler Isotrope Products France. This source will be mounted outside of the cryostat, behind an external filter wheel controlled by an arduino, which will allow us to rotate filters of varying thicknesses in front of the source in order to obtain our desired flux of $\sim$6\,keV Mn K-alpha X-rays. 

It is necessary to provide a clear optical path for the X-rays to travel from outside the cryostat to the detector, while also shielding the cryogenic detector from optical and infrared (IR) radiation, blocking electromagnetic interference, and maintaining vacuum. To accomplish this, various filters and windows will be mounted at different temperature stages of the cryostat. 

Several trades were performed to choose filter and window materials and thicknesses that provided a high level of blocking of visible and IR radiation, but that also had relatively high levels of X-ray transmission across our energy band of interest. Aluminized mylar filters (6\,$\mu$m mylar, 200\,nm Al) will be placed on the internal 50\,mK, 3\,K, and 70\,K temperature stages. For the external window at 300\,K, we use a ``LEX-HT" (Large Area X-ray Window) from Luxel that has relatively high X-ray transmission at low energies below 1\,keV and sustains a differential pressure of 1\,atm. 

\subsection{Warm Readout Electronics, Data Acquisition and Analysis}
Initially, the warm readout of the test bench microcalorimeter system will be a copy of currently-operating systems at GSFC and NIST. This will eventually be replaced by the WFEE and the DRE, with additional necessary modifications inside the cryostat. 

Data acquisition and analysis of multiplexed X-ray pulses, IV curves, spectra, etc.\,will be performed through the software program Igor and GSFC-developed Igor and Python tools. CNES is developing prototype X-IFU acquisition and analysis tools to be implemented after initial tests. 

\section{Cryostat Characterization}
 
\subsection{Magnetic Shielding Verification} \label{ssec:magfield}

The shape of the resistive TES transition is a function of the current, temperature, and magnetic field \cite{2016SPIE.9905E..2HS}. TESs (as well as SQUIDs) are thus inherently highly sensitive to static and dynamic magnetic fields. Measurements have shown that this sensitivity is larger in the direction perpendicular to the TES than parallel, by at least two orders of magnitude \cite{2013ITAS...2301505H}. For optimal performance, and deriving from X-IFU requirements on the instrument energy resolution, we place constraints on the acceptable magnetic field experienced at the detector location in the test bench. The TESs should not experience a static field any larger than 1\,$\mu$T normal to the detector plane \cite{2016RScI...87j5109B}. The maximum dynamic linear field drift during TES operation is 18\,pT/15 minutes normal to the detector plane \cite{sron2015}. Tests and optimizations of TES detector shielding have previously been performed by colleagues at GSFC to satisfy these requirements \cite{2019JLTP..194..433M}. 

During a cryostat cooldown in preparation for TES operation, the first source of magnetic field the detectors will experience is from the quasi-static terrestrial magnetic field. The magnitude of this field can range between $\sim$30--60\,$\mu$T, depending on the orientation of the detector-normal direction with respect to the Earth's magnetic field. In order to reduce this field as much as possible within the cryostat, the outer cryostat vacuum jacket has an extra shield made from mu-metal, a ferromagnetic material with high permeability.  

After this mu-metal shield is secured, the field inside the cryostat is significantly reduced, but does not go to zero. The next step in our science run procedure is to turn on the cryostat's pulse tube cooler to cool the system to $\sim$3\,K. The detector assembly, including TESs and first-stage SQUIDs, is surrounded by a superconducting Nb shield. Once this shield drops below its superconducting transition temperature, it locks in the ambient field and blocks any additional field applied externally to the shield. In order to compensate for this trapped magnetic flux, a tunable magnetic field coil on the FPA can be used to apply a nulling magnetic field to the TESs during operation. 

Next, we lower the temperature of our cryostat down to TES operating temperature by ramping up the current through the superconducting magnet coil, then lowering to a particular set point to achieve our desired cold stage temperature. In order to maintain this temperature during a typical detector operation run, the ADR magnet current ramps down slowly from this set point to zero. In order to shield the magnetically sensitive parts from this dynamic field, the ADR is surrounded by cryoperm and hiperco shields. With this shielding, the dynamic field experienced at the TES location during a run goes from $\sim$2$\,\mu$T to zero.

\begin{figure}
\begin{center}
\begin{tabular}{c}
\includegraphics[height=10cm]{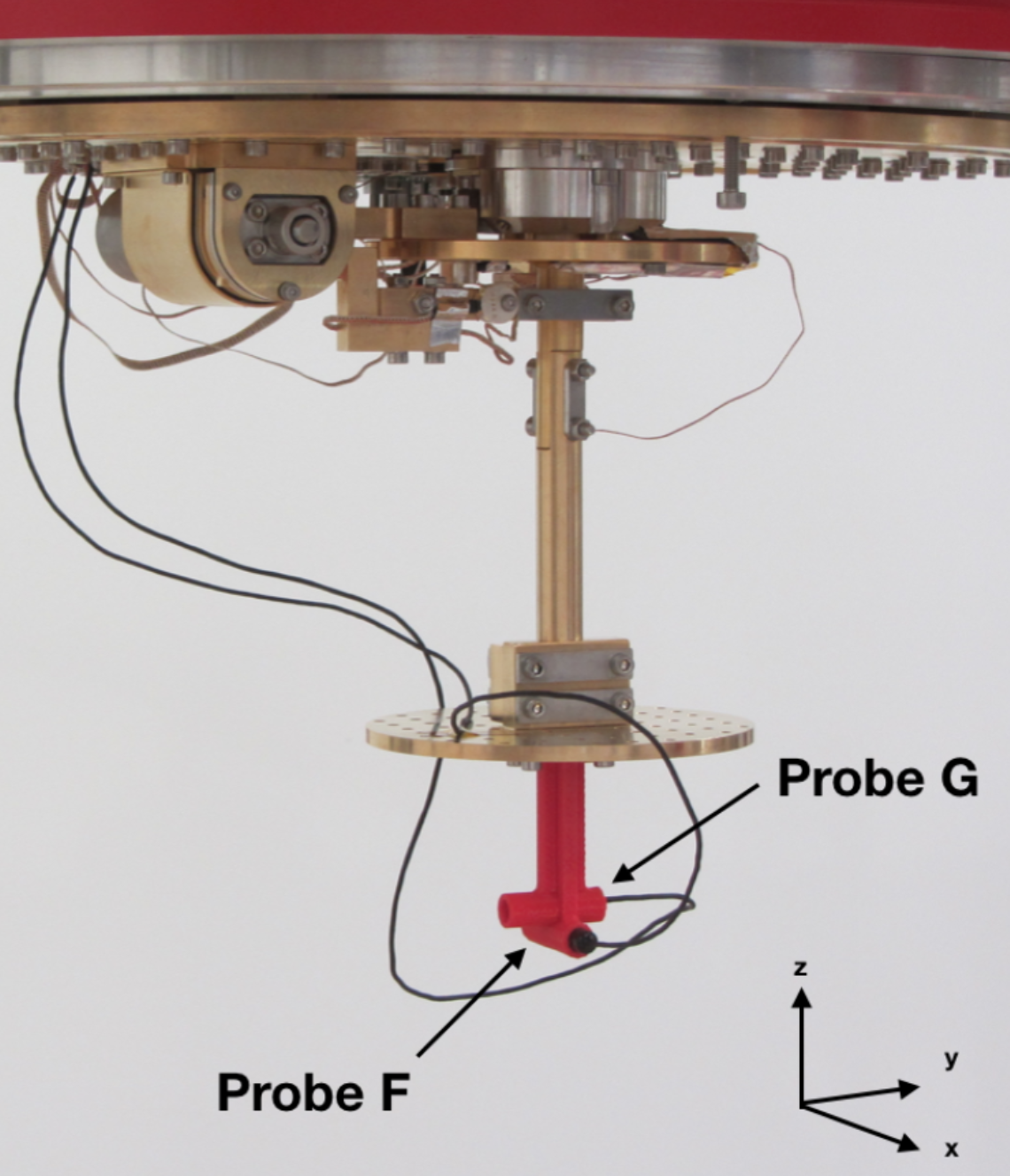}
\end{tabular}
\end{center}
\caption 
{ \label{fig:bfieldtests}
Experimental setup for magnetic fields tests within the Elsa cryostat. Red 3D printed mechanical supports are mounted onto the 50\,mK plate. The supports are holding two Bartington Instruments magnetometers pointed in orthogonal directions.} 
\end{figure}

In order to validate the shielding factors of the mu-metal and ADR magnetic shields, cryogenic Fluxgate magnetometers from Bartington Instruments were installed within orthogonally-oriented 3D-printed mounts at the nominal detector location; see Fig. \ref{fig:bfieldtests}. Since these tests were performed in 2019, this location is that of the FDM setup. The specific differences in geometry and its effects on the measured magnetic field will be described at the end of this section. 

With no mu-metal shielding, the static magnetic field from the Earth was measured to have a norm of 42\,$\mu$T\,$\pm\,5\,\mu$T, with the largest component coming from the direction perpendicular to the TDM TES chip (33\,$\mu$T\,$\pm\,0.4\,\mu$T). With the mu-metal shielding, this was reduced to a norm of 1.4\,$\mu$T\,$\pm\,0.14\,\mu$T, with a detector-normal component of 0.985\,$\mu$T\,$\pm\,0.2\,\mu$T. This is thus the residual field that will need to be compensated for by the field coil in the FPA. Our reported errors are systematic, related to the variable laboratory hardware environment. 

Magnetic field measurements were also taken during an ADR cycle to verify that the dynamic field stayed within the requirement during a normal science run. The magnetic field was measured as the cryostat was maintained at 3\,K, and the current in the ADR magnet was ramped down from 6\,A to 0\,A. The measurements were taken at the location of the detectors, but without the Nb shield. Assuming the nominal setup where the TESs are operated while the ADR magnet current decreases from 0.3\,A to 0\,A during an 11-hour hold time, the drifts deduced from the measured magnetic fields measured are 2.8\,nT/15 minutes (x direction), 1.4\,nT/15 minutes (y direction), and 25\,nT/15 minutes (z direction, normal to the detector plane of the TDM FPA). The Nb shielding has a modeled attenuation factor greater than 2000, leading to drifts lower than 13\,pT/15 minutes in any direction. This thus puts us well within the dynamic field requirement. 

The change in detector location between the FDM and TDM FPAs involves both a translation and a change in orientation. The measurements described here were taken at the location of the TES on the FDM FPA. This detector position is aligned with the axis of the ADR magnet, and the detector is side-looking. In contrast, the detector position on the TDM FPA is offset from the center of the ADR magnet by 59\,mm (radial direction), and is 60\,mm farther from the center of the ADR magnet in the axial direction. The detector on the TDM FPA is down-looking. The overall translational change in position should have minimal effects on the measured terrestrial field. According to magnetic field models provided by Entropy GmbH \cite{entropy_pc}, the change in position should also decrease the dynamic field from the ADR as seen by the detector array by approximately a factor of two, as it is farther away and offset from the magnet. Overall, our measurements reported here gives us full confidence that the magnetic field experienced by the microcalorimeters in their updated position within the cryostat will meet the test bench requirements.   

\subsection{Assessment of Microvibration}

\begin{figure}
\begin{center}
\begin{tabular}{c}
\includegraphics[height=14cm]{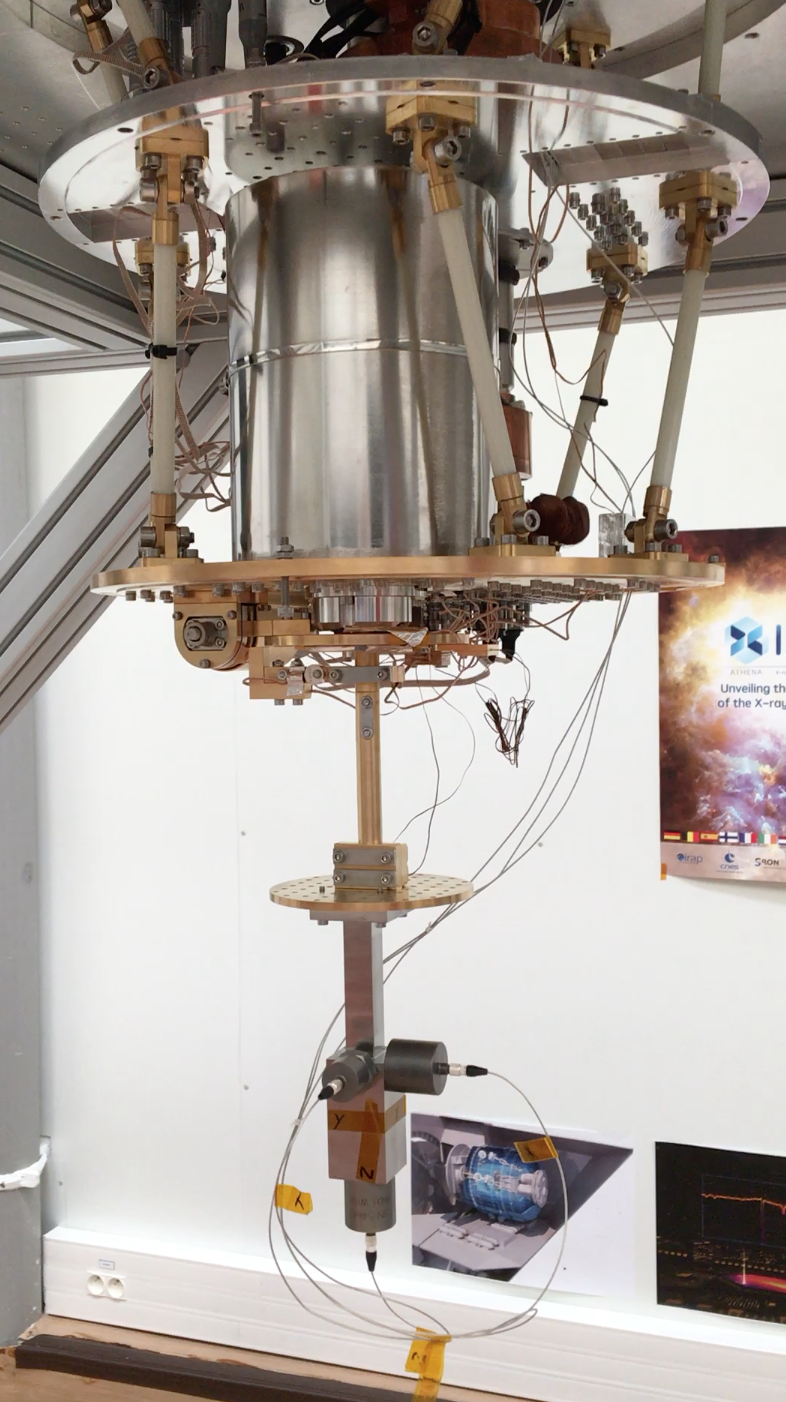}
\end{tabular}
\end{center}
\caption 
{\label{fig:accelero}
The interior of the Elsa cryostat, showing the three piezo-accelerometers mounted onto the 50\,mK cold stage.} 
\end{figure}

A source of noise that can degrade the performance of TESs is microvibration in the system, resonances of which can lead to temperature fluctuations of the thermal bath. As part of our cryostat characterization activities, we thus measured the amplitude and frequency of microvibration present in the system. Three piezo-accelerometers from PCB Piezotronics, Inc., including one cryogenic probe (model numbers 393B04 and 351B42), were installed at various x, y, and z-facing directions in the cryostat and the supporting mechanical structure in order to quantify the vibrations. These measurements were taken in various configurations: mounting the accelerometers on different locations and/or temperature shields within the cryostat, running the cryostat at different temperatures (300\,K, 2.5\,K), opening and closing the heat switch to connect the cold stages to the 3\,K stage, and turning the pulse tube on and off. A photo of the cryostat with the accelerometers mounted onto the cold stage is shown in Fig. \ref{fig:accelero}. 

The signals from the accelerometer were amplified with a signal conditioner (PCB Piezotronics, Inc. model 480E09), then acquired with a National Instruments data acquisition board (model NI 6281, M Series). The time-domain signals were then read through a LabView program, which performed a Fourier analysis to obtain a power spectral density (PSD) in frequency space. 

\begin{figure}
\begin{center}
\begin{tabular}{c}
\includegraphics[height=14cm]{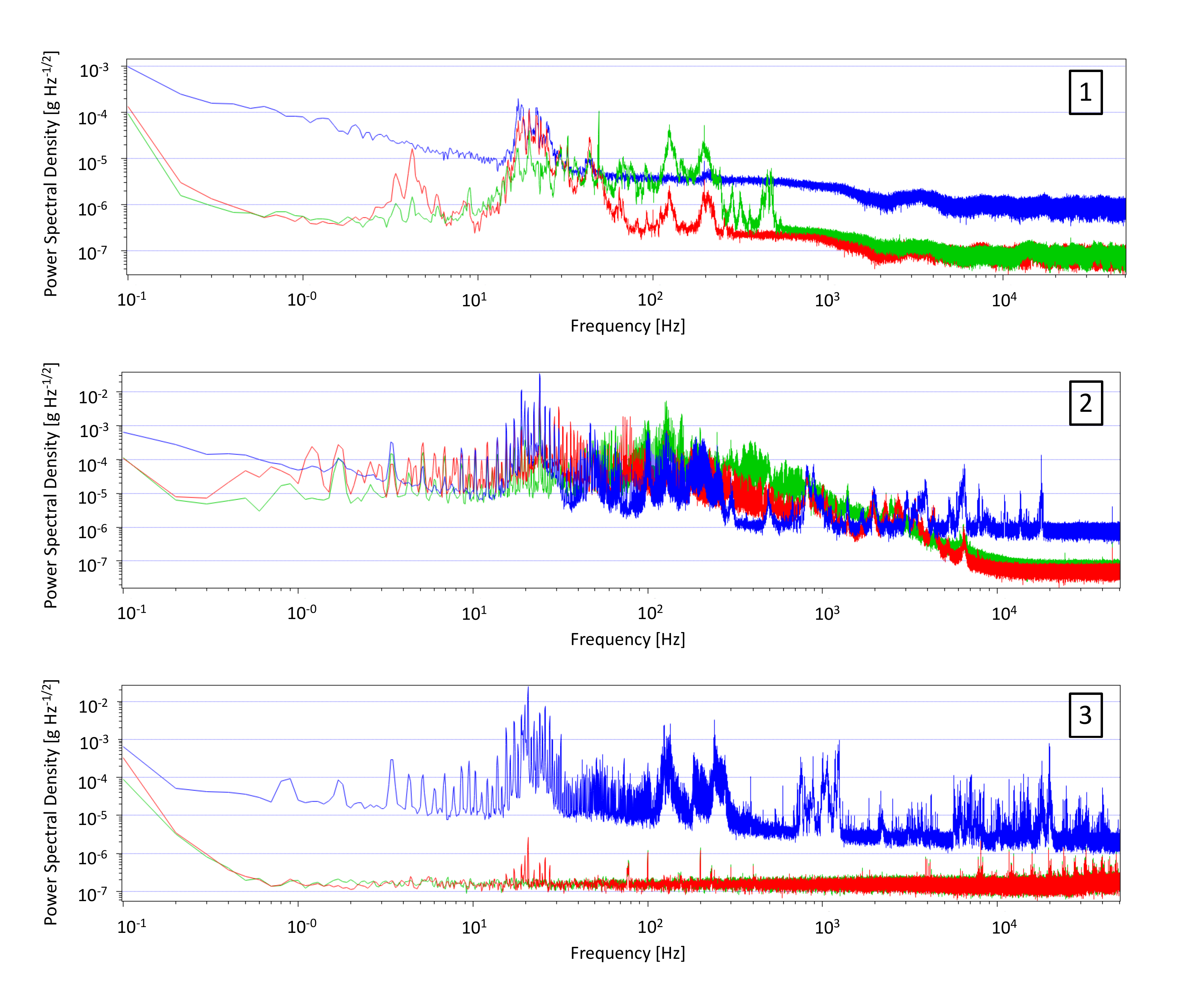}
\end{tabular}
\end{center}
\caption 
{\label{fig:psd}
Measured PSD of the microvibration signal in the x (blue), y (red), and z (green) axes in various experimental configurations: (1) 300\,K, pulse tube off, heat switch open. (2) 300\,K, pulse tube on, heat switch closed. (3) 2.5\,K at 4\,K stage, pulse tube on, heat switch open. The probe for the x direction (blue) is a cryogenic probe; this is the reason for the difference in sensitivity compared to the other probes between 300\,K and 2.5\,K.} 
\end{figure}

The measurements showed that once the pulse tube is turned on, several factors had a negligible impact on the microvibration measured, at all frequency ranges: the presence or absence of the inner thermal shields, the pressure or temperature of the cryostat, and the heat switch being open or closed. However, a peak in the PSD around 1.6\,Hz was found, along with its harmonics (see Fig. \ref{fig:psd}). It was determined that this stems from the compressor in the pulse tube cooler. There was also a broad spectral feature around $\sim$20\,Hz that is still under investigation. 

In order to mitigate these vibrations, we tested various mechanical decoupling configurations and microvibration damping strategies. These included inserting isolating material between the pulse tube and the frame of the cryostat, fully detaching the pulse tube head from the frame, and ballasting the pulse tube inlet/outlet tubes that connect to the top of the cryostat with steel pellets. However, the original, unmodified setup yielded the lowest amount of microvibration from the pulse tube at the 50\,mK stage. 

While the microvibration measurements were taken at the nominal detector location of the FDM FPA and not the TDM one (see the description of geometries in Sec. \ref{ssec:magfield}), the amplitude of the microvibrations measured at IRAP is of the same order of magnitude as that measured in cryostats at SRON\cite{sron_pc}. We thus expect that the change in the center of mass will not have a large effect on the microvibrations in the system. Further, we are confident that the level of microvibration in the system is acceptably low: temperature fluctuations of the ADR are dominated by the inelastic dissipation of the mechanical microvibrations, and as mentioned in Sec. \ref{ssec:cryostat}, we achieve a temperature stability of 3\,$\mu$K RMS at 55\,mK, which is acceptable given the requirement on the energy resolution budget for the test bench. These thermal fluctuations will induce response changes of $\sim$0.5\,eV which will add quadratically to the energy resolution. 

\section{Current Status: Installation of Microcalorimeters and Electronics}

The teams at IRAP/CNES are working with GSFC and NIST to procure, install, and test the TDM cold readout electronics and detector array. Mechanical interface parts that are to be mounted at temperature stages under 4\,K have been gold plated, and mechanical parts that will be close to the detector and SQUIDs were verified to be non-magnetic. Thermal and mechanical fit checks of GSFC/NIST parts and associated interface pieces are ongoing. All parts related to the vacuum and pumping system, electronics (including measurement devices, voltage/current sources, switch, and computers), and materials related to X-ray tests (including the Fe-55 X-ray source, cryostat window, and blocking filters) are at various stages of installation. The cryostat window has successfully undergone a leak check. Final installation and subsequent tests of the FPA are expected in summer 2021. 

In the future, the IRAP/CNES cryostat and detection chain will be used as a test platform for calibration activities for the X-IFU, including optimizing interfaces between the cryostat and calibration sources, defining calibration sequences for the engineering and flight models, and training local personnel for future calibration activities. 

%


\bibliography{biblio}   
\bibliographystyle{spiejour}   


\end{document}